\documentclass{PoS}

\title{Towards continuum limit of screening lengths with chiral Fermions
\vskip-2.3cm\hfill\small SPhT-T06/113, TIFR/TH/06-30~~~~~~~~\vskip 2cm}

\ShortTitle{Screening lengths \& chiral Fermions}

\author{\speaker{Rajiv V. Gavai} and Sourendu Gupta \\
Department of Theoretical Physics, Tata Institute of Fundamental
         Research,\\ Homi Bhabha Road, Mumbai 400005, India\\
         E-mail: \email{gavai@tifr.res.in}, \email{sgupta@tifr.res.in} }
\author{Robert Lacaze\\
         Service de Physique Th\'eorique, CEA Saclay,\\
         F-91191 Gif-sur-Yvette Cedex, France \\
         E-mail: \email{Robert.Lacaze@cea.fr} }

\abstract{We investigate mesonic screening correlators at $T=2T_c$ using the
overlap Fermions in the quenched approximation, where $T_c$ is the QCD phase
transition temperature.  Using lattices with temporal extent up to 8, we
found that both pseudoscalar and vector correlators exhibit a nice $cosh$
behaviour, leading to a plateau behaviour in the local screening masses as a
function of distance.  The $\rho$ and $\pi$ masses so determined show very
little variation with the lattice spacing $a$.  This augurs well for the use of
chiral Fermions, and further suggests the small deviations of these masses from
the ideal gas values are genuine effects of interactions.}

\FullConference{XXIVth International Symposium on Lattice Field Theory\\
                 July 23-28, 2006\\
                 Tucson, Arizona, USA}

\begin{document}

\section{Introduction}

Lattice quantum chromodynamics (QCD) predicts a new phase, called
Quark-Gluon Plasma (QGP),  at high enough temperatures.  Furthermore, it
has contributed substantially in our understanding of this phase, providing
most of the solid information we currently have about QGP.  Nevertheless,
several outstanding questions about the nature of QGP still remain.  Since
the very early days, it has been recognised that the bulk thermodynamic
quantities, such as the energy density, cannot be explained by a
straightforward weak coupling expansion.  With more precise computations on
the lattice as well as higher orders of perturbation theory, the problem
became more and more acute.   Modifications in form of certain
resummations, non-perturbative schemes, and intuitive models were proposed
to explain the discrepancy.  Quark number susceptibilities provided an
independent, mostly successful, check on these, giving the weak coupling
picture a boost, at least for $T \ge 3$-5 $T_c$.  On the other hand, the
$J/\psi$ and $\eta_c$ mesons seem to survive up to a few $T_c$.  This
together with the results from the relativistic heavy ion collider at BNL,
which suggest a very small viscosity, gave rise the picture of a strongly
coupled QGP just above $T_c$.  Recently, a strong evidence from lattice QCD
simulations emerged \cite{cbs}, suggesting that the fermionic excitations of QGP
behave like quarks already close to $T_c$. Thus an excitation carrying
unit strangeness and 1/3 baryon number also carries electric charge of
- 1/3, i.e., behaves like a strange quark.   

Clearly for a complete picture to emerge, it seems prudent to add as many
pieces of information as can be obtained.  Static screening lengths, which we
focus on in this work, constitute one such important clue.  One studies the
screening of currents in a medium to extract long-range information on its
composition.  For exciting mesons with specific quantum numbers from the
vacuum, simplest forms of currents with those quantum numbers are chosen.  They
should exhibit deconfinement related changes above the QCD phase transition
temperature ($T_c$) \cite{detar}, while yielding the known spectrum at low
temperatures. Detailed studies have shown that this indeed does happen in the
vector, and axial-vector channels: the screening above $T_c$ appears to be due
to nearly non-interacting quark anti-quark pairs in the medium \cite{mtc,tifr}.
However, the scalar and pseudo-scalar screening masses show more complicated
behavior--- strong deviations from the ideal Fermi gas, and a strong
temperature dependence.  This puzzling behavior is generic--- it has been seen
in quenched \cite{quenched} and dynamical simulations with two \cite{nf2} and
four flavors \cite{detar,mtc,tifr,nf4} of staggered quarks, as well as with
Wilson quarks \cite{wilson}.  Staggered fermions have broken flavour symmetry
on the lattice and, indeed, extrapolation to continuum limit in the quenched
case gave hints \cite{tifr2} of the deviations being cut-off effects.  However,
the extraction of these masses both for the staggered and the Wilson case was
difficult due to the complex behaviour of the correlators, which lead to the
failure of the expected plateau in the effective distance dependent masses.

Since the number of pions and their nature is intimately related to the
actually realized chiral symmetry on the lattice, one expects any good
realization of chiral Fermions on the lattice to provide better insight into
the problem of screening lengths.  Overlap Fermions \cite{neu} have the
advantage of preserving exact chiral symmetry on the lattice, i.e, at finite
lattice spacing, for any number of flavours in contrast to the Wilson fermions
which break all chiral symmetries or staggered Fermions which do so only
partially but at the expense of breaking flavour symmetry.  In our earlier work
\cite{ggl}, we showed the pion screening mass to be closer to the ideal gas
value but still distinctly lower than rho screening mass.  We employed then
lattices with 4 temporal sites, $N_t=4$, and a conjugate gradient based method
for the overlap-Dirac operator. For temperatures $1.25 \le T/T_c \le 2 $, we
observed a very mild temperature dependence of the ratio of screening mass and
the temperature.  In the current study, we address the issue of continuum limit
of these screening masses by extending to $N_t = $ 6 and 8.  We also used
lattices with long $z$ extent to investigate more carefully the lowest lying
masses and the faster Zolotarev method for the overlap-Dirac operator.  All
computations were done at $T=2T_c$, since any temperature in the range 
1.25$T_c$ to 2 $T_c$ could suffice in view of our earlier observation on the
$N_t = 4$ lattice.

\section{Simulation Details}

The massless overlap Dirac operator ($D$) can be defined \cite{neu2} in
terms of the Wilson-Dirac operator ($D_w$) for negative mass by the relation
\begin{equation}
   D = 1 + D_w (D_w^\dag D_w)^{-1/2}.
\label{overlap}\end{equation}
As in \cite{ggl}, we chose the negative mass in $D_w$ to 1.8 in this work as
well. The corresponding operator for massive quarks is 
\begin{equation}
   D(ma) = ma + (1-ma/2) D,
\label{moverlap}\end{equation}
where $m$ is the bare quark mass, $a$ the lattice spacing, and $D$
is defined above.  We used the usual quark propagator, $G(ma)=[1-D/2] 
D^{-1}(ma)$.  To compute this, one needs the inverse of the massive overlap
$D$, of eq. (\ref{moverlap}), which, as is widely known, needs a nested 
series of two matrix inversions.  At each step in the numerical inversion of 
$D$, one has to invert $(D_w^\dag D_w)^{1/2}$.  For the matrix 
$M=D_w^\dag D_w$, and a given source vector $b$, we computed
$y=M^{-1/2} b$ by using the Zolotarev algorithm \cite{wupp}:  
\begin{equation} 
M^{-1/2} b =\sum_{l=1}^{N_{\cal O}} \left(\frac{c_l}{M+d_l} b \right)~.~ 
\label{za}
\end{equation}
\noindent Here the coefficients $c_l$ and $d_l$ are computed with Jacobi
elliptic functions for a chosen  order of approximation ${N_{\cal O}}$ and a
ratio $\kappa=\mu_{\max}/\mu_{\min}$ where $\mu_{\max}$ and $\mu_{\min}$ bound
of the domain for which we apply the approximation.  In our implementation of
the algorithm \cite{us2}, we first compute the lowest and highest eigenvalues of
$M$ for a given required precision $\epsilon$.  Then we define the bound of the
domain by increasing by 10\% the domain of eigenvalues.  The order ${N_{\cal
O}}$ is defined by requiring  a precision $\epsilon/2$ for the approximation of
$1/\sqrt{z}$ in the entire domain.  We used $\epsilon = 10^{-5}$ and found that
typically ${N_{\cal O}} \sim 7-8$ was needed.  With these parameters, one
calculates the approximation in eq. (\ref{za}) by a multishift CG-inversion at
the precision $\epsilon/2$.

The lattice sizes we employed were $4 \times 10^2 \times 16$, $6 \times 14^2
\times 24$ and $8 \times 18^2 \times 32$, which ensure that the transverse
dimensions remains in the confined phase at $2T_c$. The corresponding
couplings are respectively the known critical couplings on $N_t= 8$, 12 and
16 lattices.  These $\beta$ values are 6.0625, 6.3384 and 6.55.  For the
last value, no infinite volume extrapolation was available, unlike the
first two. Our choice was motivated by the 2-loop $\beta$-function, and
consistency with the the finite volume results. In each case,
configurations separated by 1000 sweeps of a Cabbibo-Marinari update were
generated and 20-25 such configurations were used to compute the quark
propagators.  The propagator $G$ was computed on 12 point sources (3 colors
and 4 spins) for 8 quark masses from $m/T$=0.008 to 0.8 using a
multi-mass inversion of $D^{\dagger}D$.  The tolerance used was
$\epsilon=10^{-3}$ in this outer conjugate gradient.

\section{Results}

\begin{figure}[htb]
\includegraphics[scale=0.6]{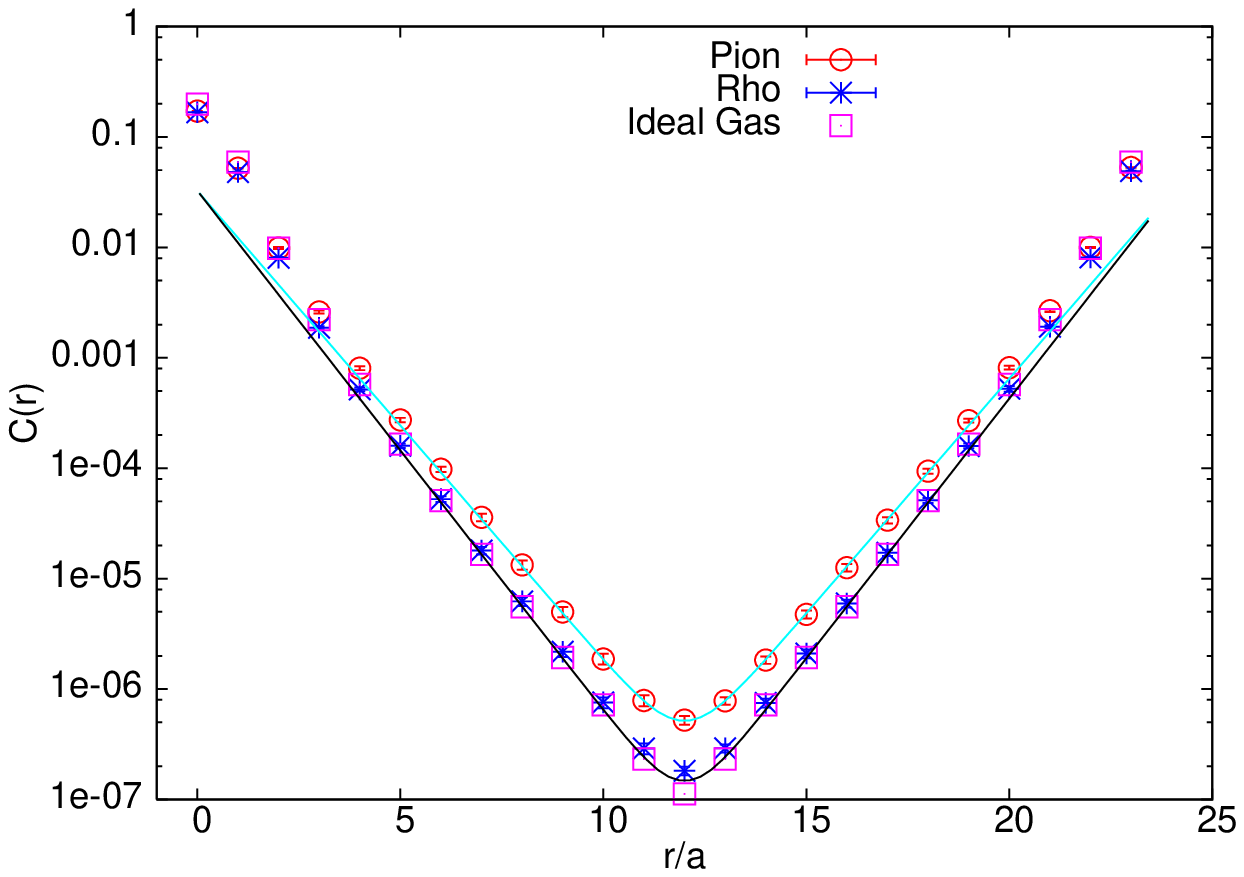}
\includegraphics[scale=0.6]{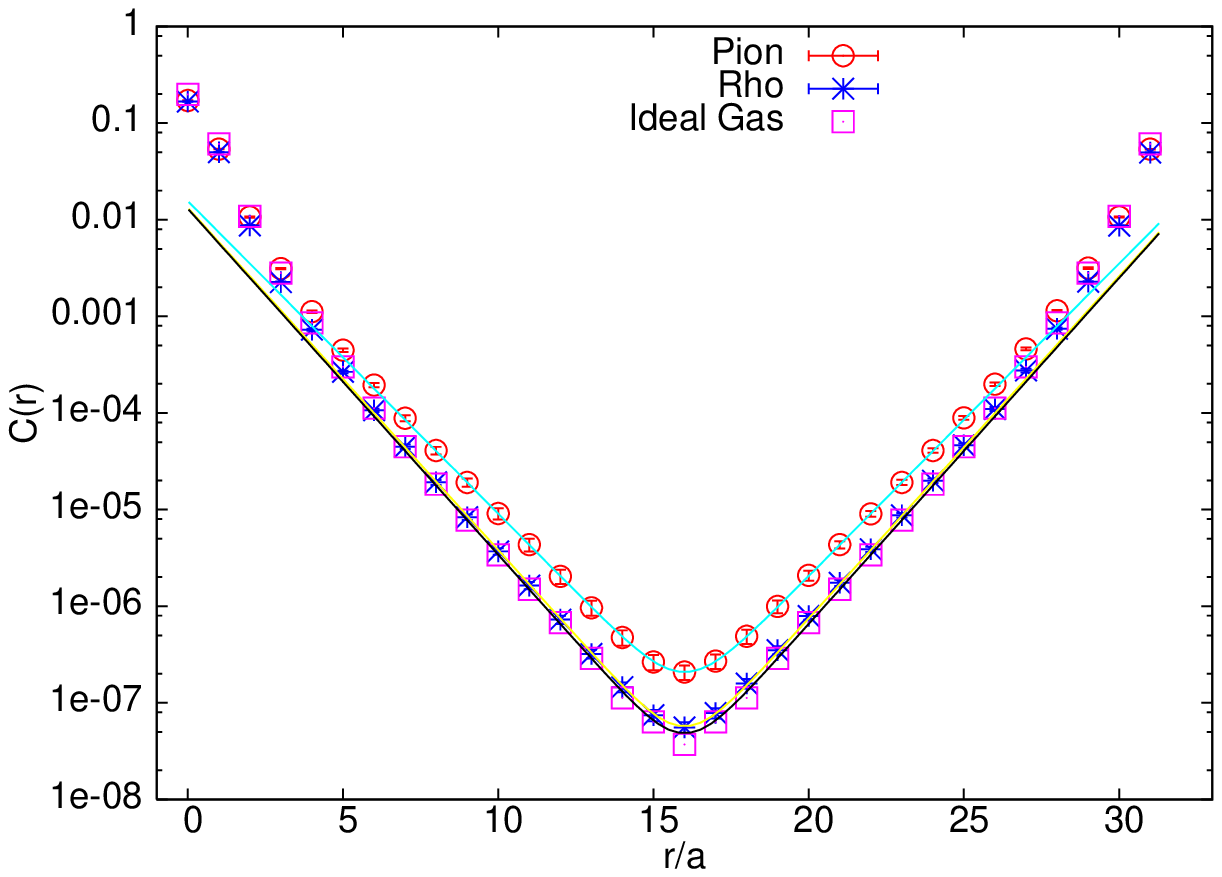}
  \caption{Correlators for pion and rho compared with the corresponding
    ideal gas results on a $6\times14^2 \times 24$ lattice (left)
    and a $8\times18^2 \times 32$ lattice (right) at $T=2 \ T_c$. 
    The lines are single $cosh$ fits.  }
\label{fg.corr}
\end{figure}

\begin{figure}[htb]
\includegraphics[scale=0.6]{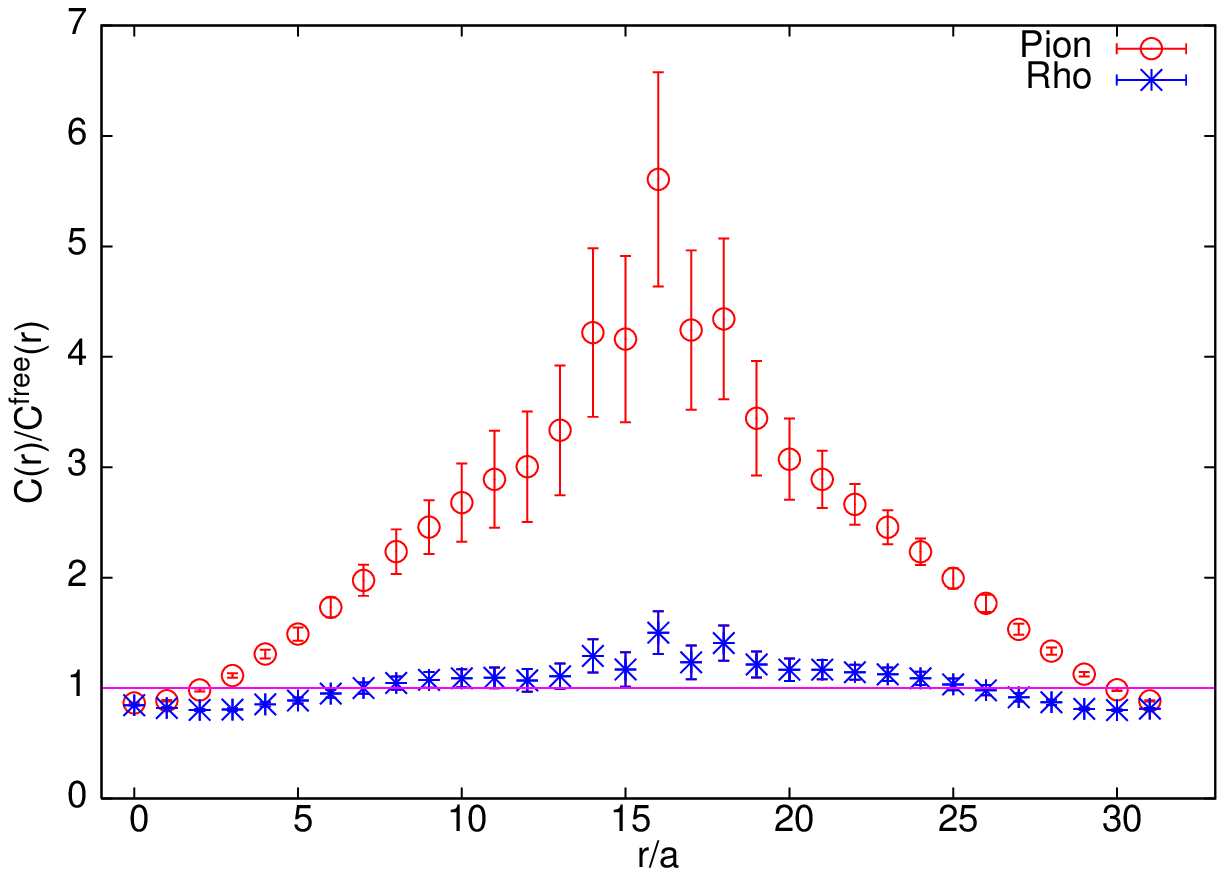}
\includegraphics[scale=0.6]{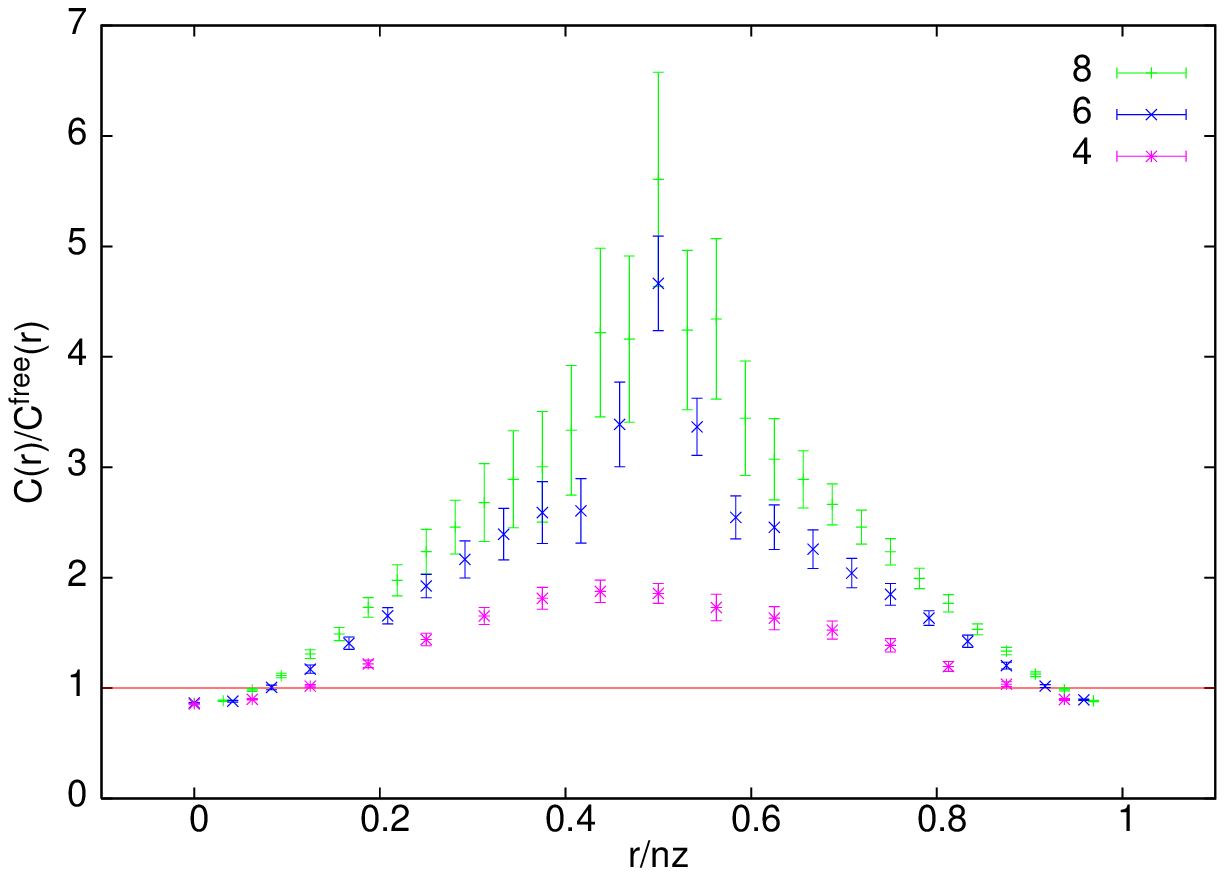}
  \caption{Ratios of pion and rho correlators on $8\times18^2 \times 32$ 
   lattice with the corresponding ideal gas results (left) and the variation of
   the pion-ratio with $N_t$  (right).   }
\label{fg.rat}
\end{figure}

Figure \ref{fg.corr} displays our results for the correlation functions of
$\rho$ and $\pi$ along with the corresponding ideal gas result on the
$6\times14^2 \times 24$ lattice (left) and  $8\times18^2 \times 32$ lattice
(right).   Both display an excellent {\em cosh} behaviour, indicative of a
dominance of a single mass scale, as the line of a {\em cosh}-fit in each
case shows.   Moreover, one sees a very good agreement of the
$\rho$-correlator with the ideal gas on both lattices in essentially its
entire 6-7 orders of magnitude fall.  There are, however, differences for
the pion.   In order to exhibit the comparison with ideal gas more clearly,

\noindent we show in the left panel of Figure \ref{fg.rat} the ratios of the
two correlators with that for ideal gas on the our largest lattice.  While
the deviations from ideal gas seem at $\sim$20 \% level for $\rho$, those
for $\pi$-correlator range up to a factor of 5-6.    Note that in both
cases the ideal gas is {\em larger} at small distances and {\em smaller} at
large distances.  As the right panel of the Figure \ref{fg.rat} shows, the
deviations 

\noindent from ideal gas appear to increase progressively in the
continuum limit, i.e, with increasing $N_t$.  Similar trend can also be
seen in the case of $\rho$, although the size of deviations and the
errorbars make it much less conclusive.  We define local masses, as usual, 
from the ratios of correlation function at successive $r$ values.
As may be anticipated from Figure \ref{fg.corr}, these local masses 

\begin{figure}[htb]
\includegraphics[scale=0.6]{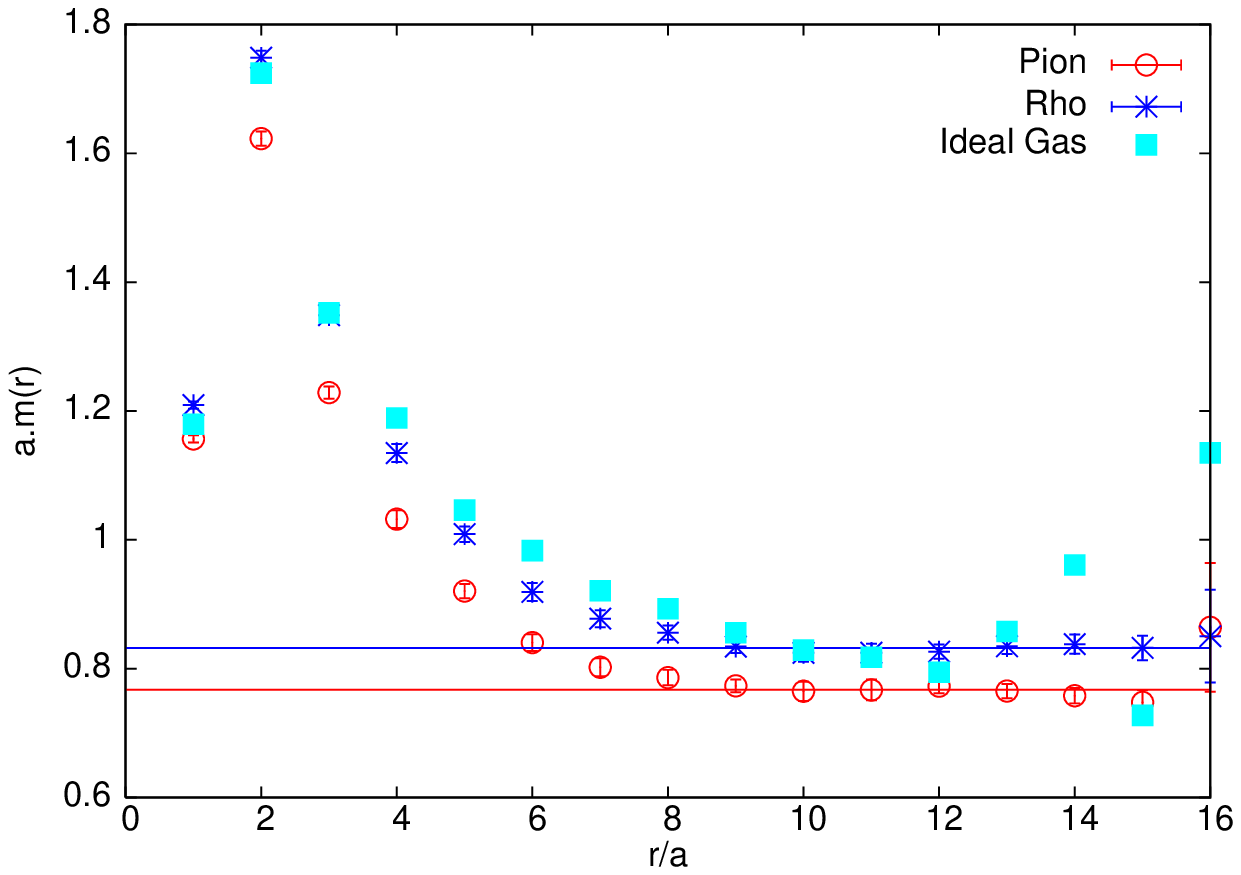}
\includegraphics[scale=0.6]{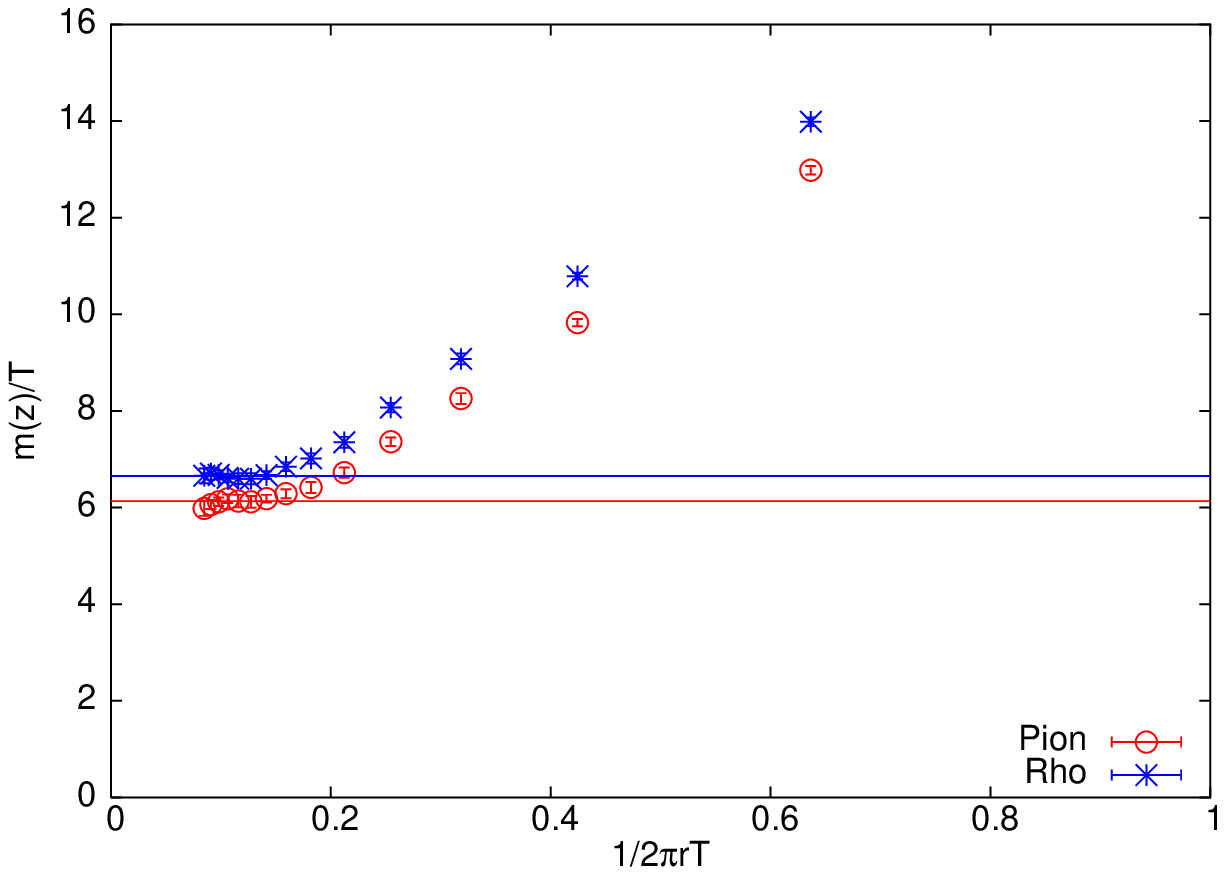}
   \caption{Left panel shows the effective mass as a function of the 
      separation $r$ on $8\times18^2\times 32 $ lattice. The horizontal lines
      indicate the corresponding mass estimate.  The right panel displays
      the plateau behaviour in another variable advocated in 
      \cite{wil05}.}
\label{fg.efz}
\end{figure}

\noindent 
for both $\rho$ and $\pi$ show demonstrate an excellent plateau behaviour.
This is exhibited in the left panel of Figure \ref{fg.efz}.  The ideal gas
results are also shown on the same panel for a comparison.  One sees very
similar behaviour for them as the $\rho$-local mass  but for the last three
points which have a distinct oscillatory pattern.   This should be contrasted
with the corresponding results for the staggered Fermions \cite{tifr2} or
Wilson Fermions \cite{wil05}, both of which display rather limited range of
plateau, if at all.  The right panel shows this explicitly by displaying the
local masses as a function of a variable, $1/2\pi rT$, advocated in
\cite{wil05}.  While the plateau behaviour is manifestly seen even in this
variable in our overlap Fermion results in the right panel of Figure
\ref{fg.efz}, no such plateau is visible in the corresponding figure 
(Figure 1) of \cite{wil05}.

The plateau in local masses also brings out the complementary nature of the
long-range information they contain.  The small-distance behaviour of the
correlator, and the corresponding local masses, have nothing to do with the
plateau value.  On the other hand, local quantities, such as the chiral
condensate, are related to the integral of the correlator and are thus
dominated by the small-distance behaviour only.  From Figure \ref{fg.corr}, one
thus surmises that the chiral condensate agrees with its ideal gas value well
on both $N_t =$ 6 and 8, although the local mass for pion in either case does
not.  Furthermore, the existence of the plateau suggests existence of genuine
bound states in both $\rho$ and $\pi$ channels.

\begin{figure}[htb]
\includegraphics[scale=0.6]{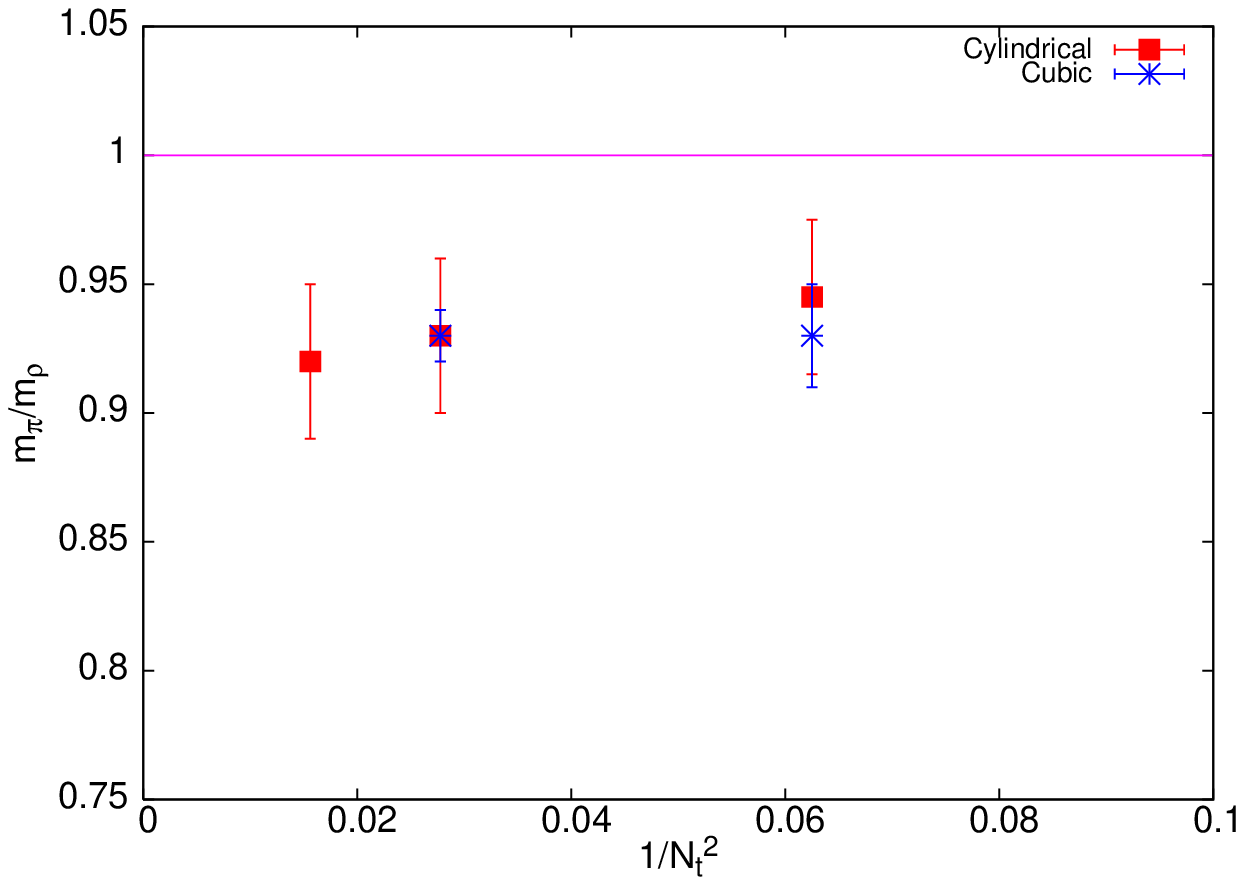}
\includegraphics[scale=0.6]{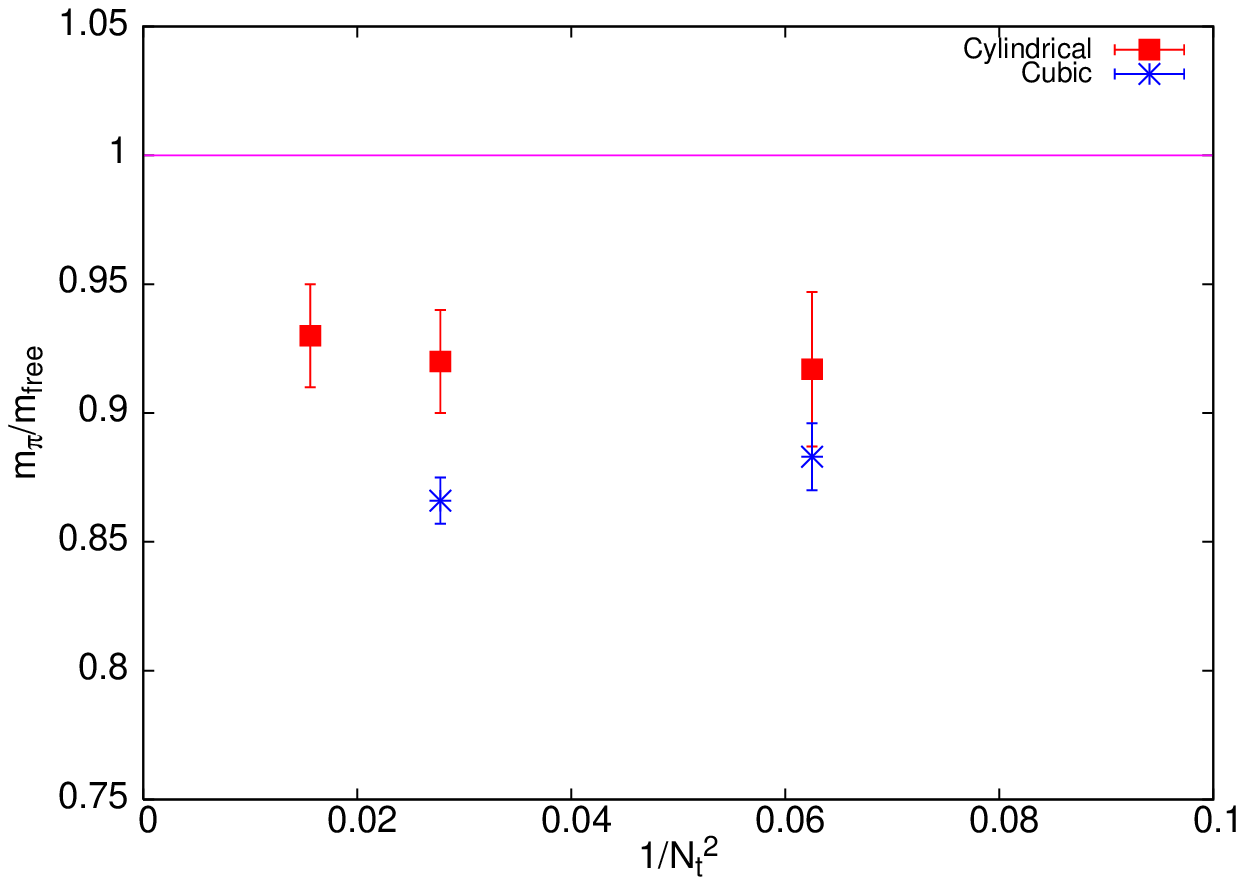}
\caption{Continuum limit of the ratios  $m_\pi/m_\rho$ (left)
      and $m_\pi/m_{free}$ (right).} \label{fg.efm}
\end{figure}

Figure \ref{fg.efm} exhibits our results on $N_t = 6$ and $8$, together
with our earlier results \cite{ggl} on $N_t = 4$ for the ratio of
$m_\pi/m_\rho$ (left) and $m_\pi/m_{free}$ (right).  The estimates for
$m_\pi$ and $m_\rho$ were obtained from the plateaux in Figure
\ref{fg.efz}, and checked independently by a direct single cosh fit to the
correlation function in the interval of the local mass plateau in each
case.  The $m_{free}$ too was obtained from a fit, and was found to agree
with the corresponding value of its smaller plateau region.  Also shown are
our results on cubic lattices having the same transverse size.  Both panels
display very little, if any, lattice spacing, $a$, dependence, suggesting
the pion screening length to remain about 8 \% below the ideal gas value,
or more than 2 $\sigma$ away, whereas the $m_\rho$ is consistent with the
ideal gas, as seen by comparison of the two panels or alternatively Figure
\ref{fg.efz}.  The ratio $m_\pi/m_\rho$ does not appear to get affected by
the geometry of the lattice, whereas the ratio $m_\pi/m_{free}$ seems
smaller on the cubic lattice.

\section{Summary}

Extending our earlier work \cite{ggl} on the hadronic screening lengths on $N_t
= 4$ lattice to $N_t = 6$ and 8 at a fixed temperature of 2$T_c$ on lattices
with large $z$ extent, we found a single {\em cosh} behaviour in both $\rho$
and $\pi$ correlators on both lattices.  This led to a convincing plateau
behaviour in the corresponding local masses, suggesting the presence of a bound
state in each case.   The $\rho$-correlator appeared to be in very good
agreement with the ideal gas correlator on the same lattice whereas the
$\pi$-correlator differed from it on all $N_t$.  In fact, the deviations appear
to {\em increase} in the continuum limit, i.e, with increasing $N_t$.  Ratios
of the extracted screening masses displayed very little, if any, dependence on
the lattice spacing.  While the $\rho$ screening mass is in agreement with the
free value, the $\pi$ screening mass remained lower by about 8 \% or more than
2$\sigma$.  Its constancy with the lattice spacing $a$ suggests the deviation
to be a genuine interaction effect.

\section{Acknowledgements}

The computations reported here were performed on the CRAY X1 of the Indian
Lattice Gauge Theory Initiative at the Tata Institute of Fundamental
Research.  It is a pleasure to thank the system administrator, Ajay Salve.
The kind hospitality of the Service de Physique Th\'eorique, Saclay, where
this manuscript was completed during a visit, is gratefully acknowledged.
This work was funded by the Indo-French Centre for the Promotion of
Advanced Research under its project number 3104-3.

\end{document}